\documentclass[twocolumn,aps,prl,showpacs,superscriptaddress,preprintnumbers,floatfix,nofootinbib]{revtex4}
\usepackage{multirow}
\usepackage{hhline}
\usepackage{bbm}
\usepackage{epsfig}
\usepackage{graphics}  % needed for figures
\usepackage{graphicx}  % needed for figur\es
\usepackage{dcolumn}   % needed for some tables
\usepackage{bm}        % for math
\usepackage{amssymb}   % for math
\usepackage{amsmath}   % for math
\usepackage{amsfonts}
\usepackage[percent]{overpic}
\usepackage{natbib}
\usepackage[colorlinks,linkcolor=blue,urlcolor=blue,citecolor=blue]{hyperref}
\usepackage[usenames]{color}
\usepackage{CJK}

\begin{document}

\begin{CJK*}{GBK}{song}

%????????????????????????????????????????????????????????????????????????????????????????????????????????????????????????
%0
\title{Giant Dipole Resonance as a Fingerprint of $\alpha$ Clustering Configurations in $^{12}$C and $^{16}$O}

\author{W. B. He}
\affiliation{Shanghai Institute of Applied Physics, Chinese Academy of Sciences, Shanghai 201800, China}
\affiliation{University of the Chinese Academy of Sciences, Beijing 100080, China}

\author{Y. G. Ma\footnote{ygma@sinap.ac.cn}}
\affiliation{Shanghai Institute of Applied Physics, Chinese Academy of Sciences, Shanghai 201800, China}
\affiliation{Shanghai Tech University, Shanghai 200031, China}

\author{X. G. Cao\footnote{caoxiguang@sinap.ac.cn}}
\affiliation{Shanghai Institute of Applied Physics, Chinese Academy of Sciences, Shanghai 201800, China}

\author{X. Z. Cai}
\affiliation{Shanghai Institute of Applied Physics, Chinese Academy of Sciences, Shanghai 201800, China}
\author{G. Q. Zhang}
\affiliation{Shanghai Institute of Applied Physics, Chinese Academy of Sciences, Shanghai 201800, China}

\date{Received 6 May 2014; published 17 July 2014}
\begin{abstract}
 It is studied  how the $\alpha$ cluster degrees of freedom, such as $\alpha$ clustering configurations close to the $\alpha$ decay threshold in $^{12}$C and $^{16}$O, including the linear chain, triangle, square, kite, and tetrahedron, affect nuclear collective vibrations with a microscopic dynamical approach, which can describe properties of nuclear ground states well across the nuclide chart and reproduce the standard giant dipole resonance (GDR) of  $^{16}$O quite nicely.
It is found that the GDR spectrum is highly fragmented into several apparent peaks due to the $\alpha$ structure.
The different $\alpha$ cluster configurations in $^{12}$C and $^{16}$O have corresponding characteristic spectra of GDR.
The number and centroid energies of peaks in the GDR spectra can be reasonably explained by the geometrical and dynamical symmetries of $\alpha$ clustering configurations.
Therefore, the GDR can be regarded as a very effective probe to diagnose the different $\alpha$ cluster configurations in light nuclei.
\end{abstract}
%\doi{10.1103/PhysRevLett.113.032506}
\maketitle

{\it Introduction.}---Clustering is one of the most fundamental physics aspects in light nuclei. It is typically observed as excited states of those nuclei and also in the ground states for nuclei far from the $\beta$ stability line, where nuclei can  behave like molecules composed of nucleonic clusters.
A great deal of research work has been focused on $\alpha$ clustering for more than four decades \cite{greiner1995nuclear,Vonoertzen2006PR432}.
It is well established that $\alpha$  clustering plays a very important role in self-conjugate light nuclei near the  $\alpha$ decay threshold due to the high stability of the $\alpha$ particle and the strong repulsive $\alpha$-$\alpha$ interaction \cite{Ikeda1968PTPSE68,Vonoertzen2006PR432,schuck2003alpha}. At low densities and temperatures, strong alpha clustering of the nuclei is also predicted \cite{Girod2013PRL111}.
An important effect on the nuclear equation of state  due to the clustering effect was also reported at low densities \cite{Ropke2013NPA897,Horowitz2006NPA776,Natowitz2010PRL104,Qin2012PRL108}.
The influence of clustering on nucleosynthesis is a fundamental problem to answer in nuclear astrophysics \cite{Holy}.
However, many problems have not yet been well understood, such as how $\alpha$ clustering determines the configurations and shapes of the many-body system and what are the aspects of  the collective dynamics of $\alpha$ clustering systems and the underlying mechanism, etc. \cite{Umar2010PRL104,Ichikawa2011PRL107,Ebran2012Nature487,FREER2007RPP70,FUNAKI2010JPG37}.

Isovector nuclear giant dipole resonances (GDRs), as the most pronounced feature in the excitation of nuclei throughout the whole nuclide chart, can give crucial clues to understand nuclear structure and collective dynamics. It is well established that the centroid energy of this resonance can provide direct information about nuclear sizes and the nuclear equation of state \cite{Harakeh2001giant}. Meanwhile, the GDR width closely relates with nuclear deformation, temperature, and angular momentum \cite{Harakeh2001giant,Pandit2010PRC81,Pandit2013PRC87}.
The GDR strength has a single peak distribution for spherical nuclei with mass number $>$ 60. The GDR in light nuclei is usually fragmented \cite{Eramzhyan1986PR136,Harakeh2001giant,Yamagata2004PRC69}.
For nuclei far from the $\beta$ stability line, another low-lying component appears called pygmy dipole resonance   \cite{Leistenschneider2001PRL86,Bacca2002PRL89,Adrich2005PRL95,Kanada-En'yo2005PRC72}, which relates with the oscillation between the valence nucleons and the core.

It can be expected that multifragmented peaks, rather than only one broad peak in the GDR spectra, can also be obtained for self-conjugate ($\alpha$) nuclei such as $^{12}$C and $^{16}$O with a prominently developed $\alpha$ cluster structure in excited states.
Therefore, it is very interesting to study how an $\alpha$ cluster component manifests itself in GDRs.
The GDR spectra should provide important and direct information to reveal the geometrical configurations and dynamical interactions among $\alpha$ clusters.
In this work, we report on the results of GDRs of $\alpha$ cluster states in light excited self-conjugate nuclei within  a microscopic dynamical many-body approach. Then, the way in which the different $\alpha$ configurations affect the GDR distributions is investigated and the underlying mechanism responsible for the collective motion is addressed.

For $^{12}$C, triangularlike configuration, is predicted around the ground state by fermionic molecular dynamics \cite{Chernykh2007PRL98}, antisymmetrized molecular dynamics \cite{Kanada-En'yo2012PTEP2012,Furuta2010PRC82}, and covariant density
functional theory \cite{Liu2012CPC36}, which is supported by a new experiment \cite{addref1}.
A three-$\alpha$  linear-chain configuration was predicted as an excited state with
different approaches \cite{Morinaga1966PL21,Umar2010PRL104,Liu2012CPC36}.
The intrinsic density of $^{12}$C and $^{16}$O may display localized linear-chain density profiles as an excitation of the condensed gaslike states described with the Brink wave function and the Tohsaki-Horiuchi-Schuck-R\"{o}pke wave function \cite{THSR2001PRL87,schuck2003alpha,Suhara2014PRL112}.
For $^{16}$O, the linear-chain structure with four-$\alpha$ clusters  was supported by the alpha cluster model \cite{Bauhoff1984PRC29} and the cranked Skyrme Hartree-Fock method \cite{Ichikawa2011PRL107}.
A tetrahedral structure of  $^{16}$O, made out of four-$\alpha$ clusters, is found above the ground state with the constrained Hartree-Fock-Bogoliubov approach \cite{Girod2013PRL111}.
However, recent calculations with chiral nuclear effective field theory \cite{Epelbaum2014PRL112}, covariant density functional theory \cite{Liu2012CPC36}, and an algebraic model \cite{Bijker2014PRL112} also support the tetrahedral $\alpha$ configuration in the ground states.
Recent orthogonality condition model calculations show a duality of the mean-field-type as well as $\alpha$-clustering character in the $^{16}$O ground state \cite{Yamada2012PRC85}.
There are also many different configurational descriptions implying the $\alpha$ cluster structure  in  $^{20}$Ne and $^{24}$Mg, such as three-dimensional shuttle shape \cite{Girod2013PRL111,Ebran2012Nature487} or chain states \cite{Chappell1995PRC51,Marsh1986PLB180} as well as nonlocalised cluster states \cite{ZHOU2013PRL110}.
Therefore, it is highly necessary and important \cite{Broniowski2014PRL112} to look for new probes to diagnose different configurations for $\alpha$-conjugate nuclei around the cluster decay threshold.

{\it Model  and methodology.}---
Quantum molecular dynamics (QMD) type  models have been successfully applied recently for the study of various giant resonances (including GDR, pygmy dipole resonance  and giant monopole resonance) due to its microscopic basis and high flexibility \cite{Kanada-En'yo2005PRC72,Furuta2010PRC82,Wu2010PRC81,Tao2013PRC87,CAO2010PRC81}.

In the following calculations of GDRs, the nuclear system is described within the QMD model framework. To apply this approach to light nuclei like $^{12}$C and $^{16}$O, two features of the model are important. One is the capacity to describe nuclear ground states. The other is the stability of nuclei in the model description. In this respect, it should be pointed out that standard QMD shows insufficient stability due to the fact that the initialized nuclei are not in their real ground states.
To solve this problem, an extended QMD called EQMD that displays some new features will be applied in this work \cite{Maruyama1996PRC53,Wada1998PLB422}.
For instance, the width of Gaussian wave packets for each nucleon is independent and  treated as a dynamical variable, which is an important improvement compared with older models with a uniform and static width for all nucleons.
Furthermore, the kinetic-energy term arising from the momentum variance of wave packets is taken into account  by subtracting the spurious zero-point center of mass (c.m.) kinetic energy from the Hamiltonian.
In standard QMD, the kinetic-energy term arising from the momentum variance of wave packets is spurious.
Thus, the constituent nucleons having finite momenta  are not in energy-minimum states that are the source of insufficient stability.

For the effective interaction, Skyrme and Coulomb forces, as well as symmetry energy  and the Pauli potential, are used.
Specifically, the Pauli potential is written as
\begin{eqnarray}
H_{Pauli} = \frac{c_P}{2}\sum_i(f_i-f_0)^\mu\theta(f_i-f_0),\\
f_i\equiv\sum_j\delta(S_i,S_j)\delta(T_i,T_j)|\langle\phi_i|\phi_j\rangle|^2,
\end{eqnarray}
where $f_i$ is the overlap of a nucleon $i$ with nucleons having same spin and isospin and $\theta$ is the unit step function. The coefficient $c_P$ is the strength of the Pauli potential. This potential inhibits the system to collapse into the Pauli-blocked state at low energy  and gives the model capability to describe $\alpha$ particle clustering. This capability is very important for our calculation because it gives us the possibility to extract information about clustering configurations from GDR spectra. The phase space of nucleons is obtained initially from a random configuration. To get the energy-minimum state as ground state as a ground state, a frictional cooling method is used for the initialization process. The model can describe the ground state properties, such as binding energy, rms radius, and deformation etc., quite well over very a wide mass range.

The macroscopic description of GDRs by the Goldhaber-Teller model \cite{Goldhaber1948PR74}, which assumes that protons and neutrons collectively oscillate with opposite phases in an excited nucleus, is used to calculate it from the nuclear  phase space obtained from the EQMD model.
Specifically, we get the initial state wave function $\Psi(0)$ of the system, by the EQMD initialization process. Then, we boost $\Psi(0)$ at $t$ = 0 fm/$c$ by imposing a dipole excitation:
\begin{equation}
|\Psi^{*}\big>=\hat{D}(E)|\Psi(0)\big>,
\end{equation}
\begin{equation}
\hat{D}(E)=\prod_{j}\exp(-iT_j\delta{\bf x}\cdot{\bf p}_j/\hbar).
\end{equation}
Here, $\hat{D}(E)$ is the dipole excitation operator and $E$ is the excitation energy, which can be obtained by calculating the difference in energy using the  $\Psi(0)$ and $\Psi^{*}$ states. $T_j$ stands for the isospin of the nucleons , $\delta{\bf x}$ is the separation between the c.m. of neutrons and the c.m. of protons by boost.

The evolution of the excited wave function to the final state is obtained by the EQMD model, a detailed process of which was described  in Ref. \cite{Maruyama1996PRC53}.
Considering the lifetime of GDR excitations, we take the final state at $t$ = 300 fm/$c$.

The dipole moments of the system in coordinate space $D_{G}(t)$ and momentum space $K_{G}(t)$ are, respectively, defined as follows \cite{Baran2001NPA679,Wu2010PRC81,Tao2013PRC87}:
\begin{eqnarray}
D_{G}(t) = \frac{NZ}A\bigg[R_Z(t)-R_N(t)\bigg],\\
K_{G}(t) = \frac{NZ}{A\hbar}\bigg[\frac{P_Z(t)}Z-\frac{P_N(t)}N\bigg],
\end{eqnarray}
where $R_Z(t)[P_Z(t)]$ and $R_N(t)[P_N(t)]$ are the c.m.'s of the protons and neutrons in coordinate (momentum) space, respectively. $K_{G}(t)$ is the canonically conjugate momentum of $D_{G}(t)$.

From the Fourier transform of the second derivative of $D_{G}(t)$ with respect to time, i.e.,
\begin{equation}
D^{''}(\omega) = \int_{t_0}^{t_{max}}D^{''}_G(t)e^{i\omega t}dt,
\end{equation}
the strength of the dipole resonance of the system at excited energy $E = \hbar \omega$ can be obtained, i.e.,
\begin{equation}
\frac{dP}{dE} = \frac{2e^2}{3\pi\hbar c^3E}\big|D^{''}(\omega)\big|^2,
\end{equation}
where $dP/dE$ can be interpreted as the nuclear photoabsorption cross section. It can be normalized as
$(dP/dE)_{norm} = (dP/dE) \Delta E/\int_0^\infty (dP/dE) dE$, where $\Delta E$ is the energy range of the GDR concerned.
In realistic calculations, we take the integral interval from 8 to 40 MeV, which is consistent with the energy region of the GDR. The normalized $dP/dE$ is calculated in the excitation-energy region from 8 to 35 MeV, which includes almost all the physically relevant GDR peaks.
When displaying the $dP/dE$ spectrum, a smoothing parameter $\Gamma = $ 2 MeV was used (our calculation shows that the GDR width
almost does not depend on $\Gamma$).

\begin{figure}[htbp]
\flushleft
    \begin{overpic}[width=0.5\linewidth]{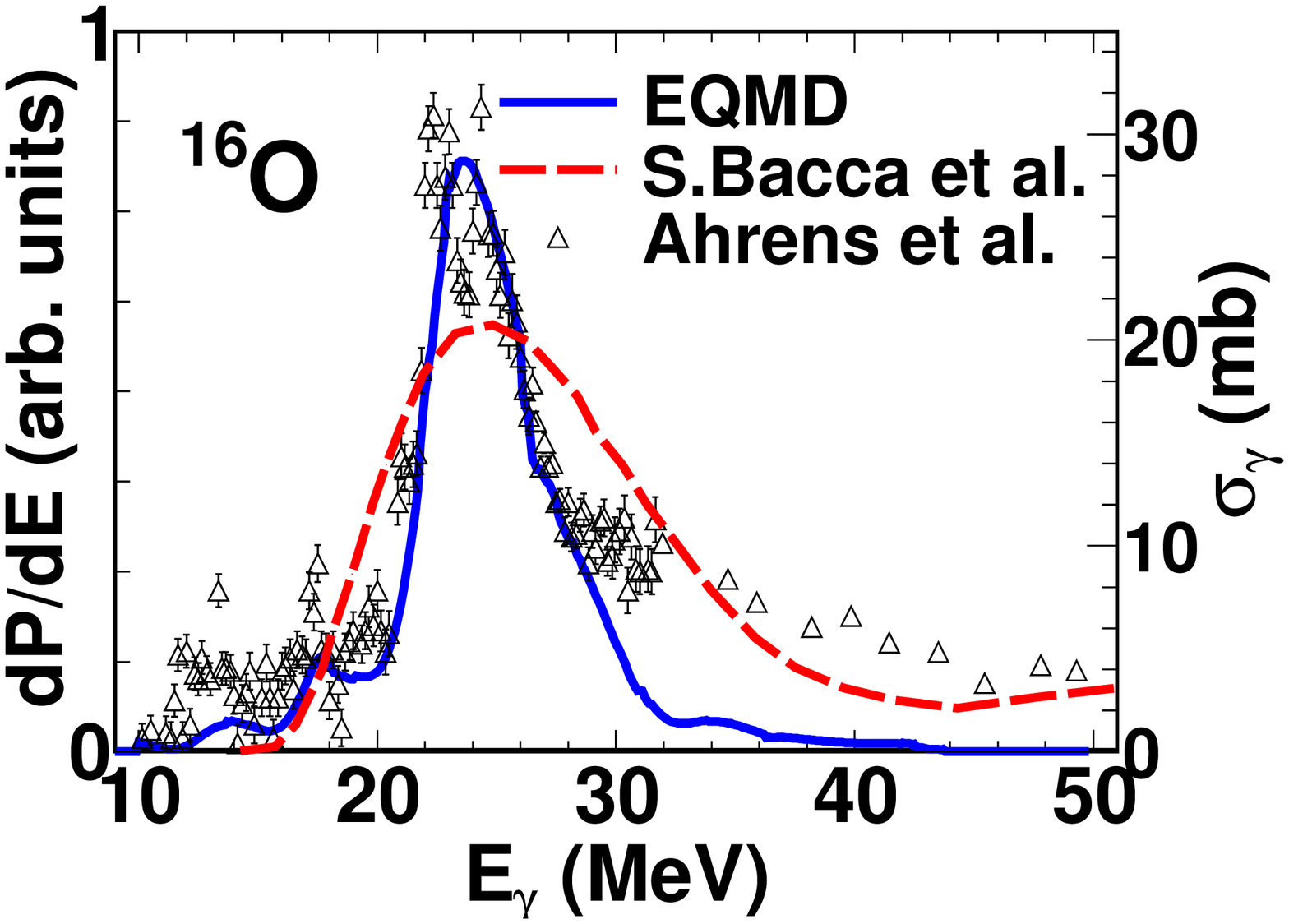}
        \put(100,0){\includegraphics[width=.5\linewidth]{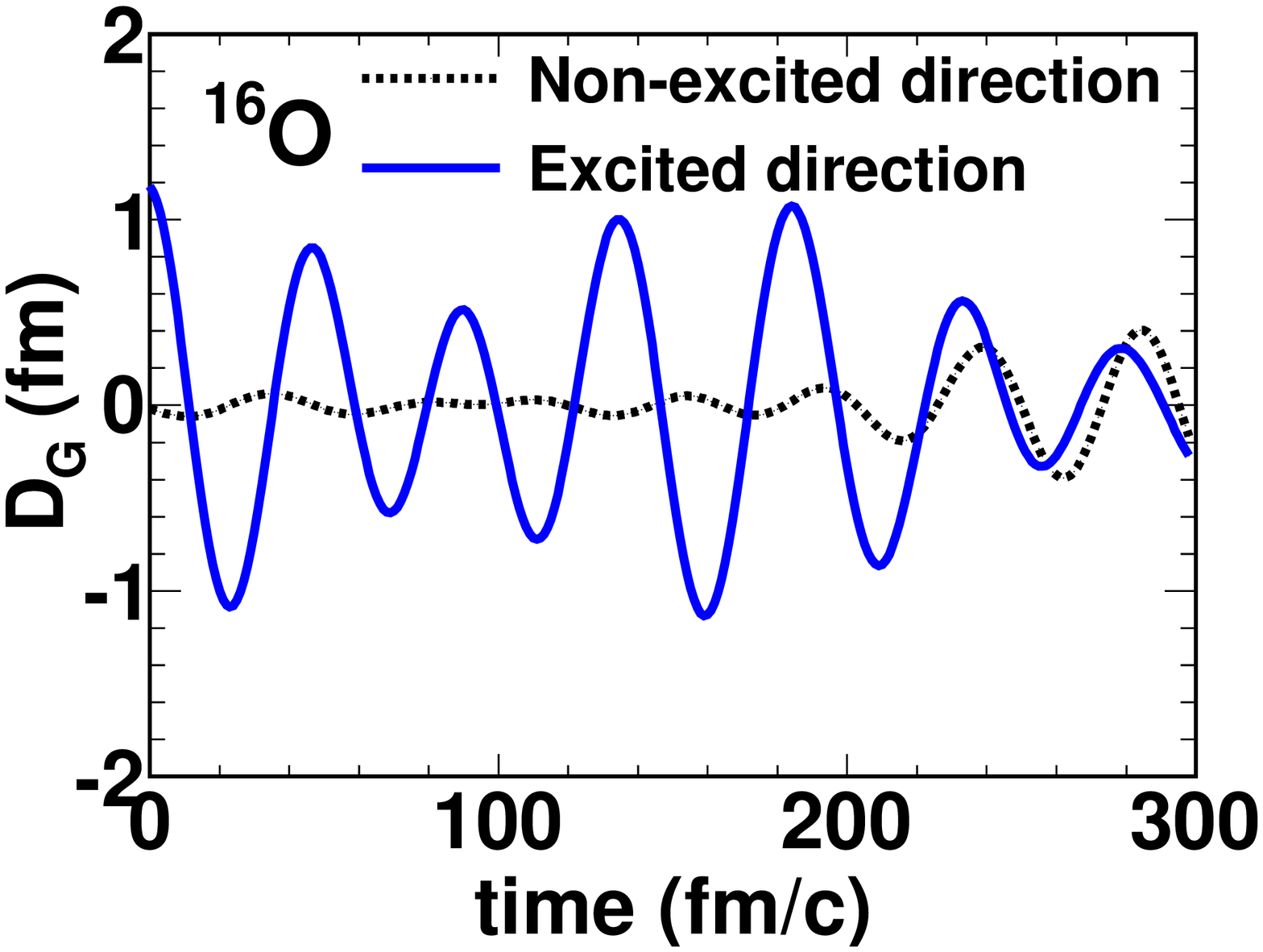}}
        \put(70,25){\bf (a)}
        \put(175,25){\bf (b)}
    \end{overpic}

\caption{\footnotesize(color online)(a) Comparison of the GDR calculation for $^{16}$O (solid blue line, scaled by the left $Y$ axis) against experimental data (nuclear photoabsorption cross section on the Oxygen target), Ref. \cite{Ahrens1975NPA251}  (empty triangles, scaled by the right $Y$ axis), and first principles calculation \cite{Bacca2013PRL111}  (long dashed red line, scaled by the right $Y$ axis),
(b) Time evolution of $D_{G}$ in the excited direction (solid blue line) and the nonexcited direction (short dashed black line).}
\label{fig:o16_ground}
\end{figure}

{\it Results and discussion.}---
The GDR spectrum of $^{16}$O obtained in the way described above is compared against the experimental data \cite{Ahrens1975NPA251} and first principle calculations \cite{Bacca2013PRL111} shown in Fig. \ref{fig:o16_ground}(a). Figure. \ref{fig:o16_ground}(b) shows the $^{16}$O dipole oscillation in two decomposed directions versus time for one event.
The wave function of the $^{16}$O system at the ground state is obtained at a binding energy of 7.82$A$ MeV, which is very close to the experimental binding energy:  7.98$A$ MeV.
The resulting ground state consists of four $\alpha$ particles with a tetrahedral configuration. The tetrahedral four-$\alpha$ configuration in the $^{16}$O ground state is also supported by a new $ab initio$ calculation by Epelbaum $et~al.$ \cite{Epelbaum2014PRL112} using chiral nuclear effective field theory. In addition, a recent covariant density
functional theory calculation also shows regular tetrahedral four-$\alpha$ configuration in the ground state of $^{16}$O \cite{Liu2012CPC36}.
The long dashed red line represents the calculated GDR of $^{16}$O by a merged Lorentz integral transform of a dipole response function obtained with the coupled-cluster method from first principles.
The comparison with data confirms that the tetrahedral four-$\alpha$ configuration in initialization is reasonable and the procedure used to calculate GDRs is reliable. Then, we apply the method to explore GDRs for excited $\alpha$ cluster states.

For light stable nuclei, the $\alpha$ cluster structure is expected around the threshold energy $E_{n\alpha}^{thr} = nE_{\alpha}$ of the $n\alpha$ emission. The Pauli principle plays a more and more important role when the $\alpha$ cluster degrees of freedom become more pronounced. Therefore, to quantitatively depict the energy of $\alpha$ cluster states, the running parameter of $c_P$, which  depends on the density, excitation energy, or temperature of the system, is needed. Thus, the $\alpha$ clustering states with different configurations around the threshold $E_{n\alpha}^{thr}$ are obtained with 20 MeV Pauli potential strength, where $\alpha$ clusters are weakly bound, less than 1 MeV per cluster, in all systems considered.

\begin{figure}[htbp]%GDR
    \centering
    \begin{overpic}[width=\linewidth]{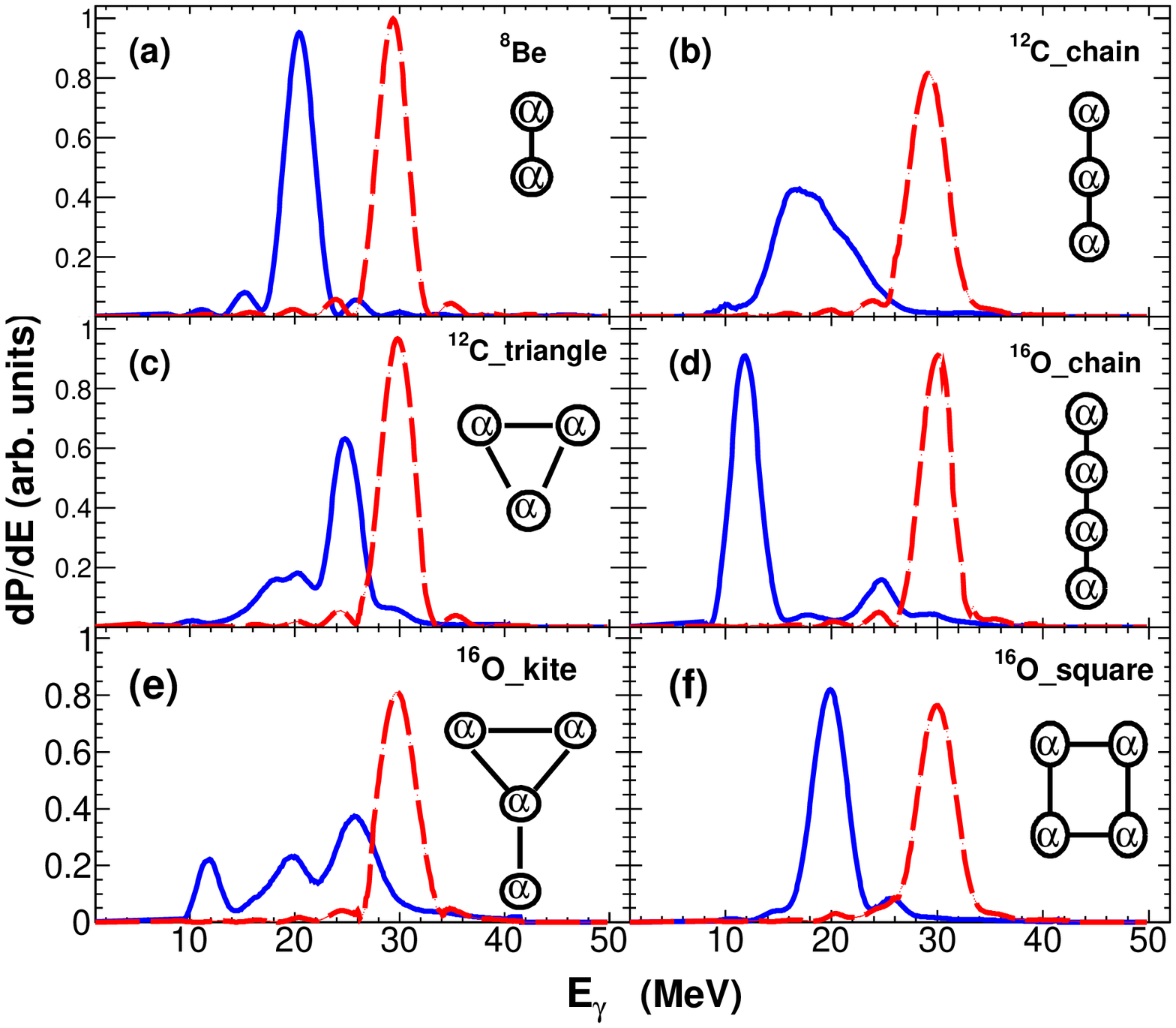}
           \put(42.7,65.5){\includegraphics[scale=0.12]{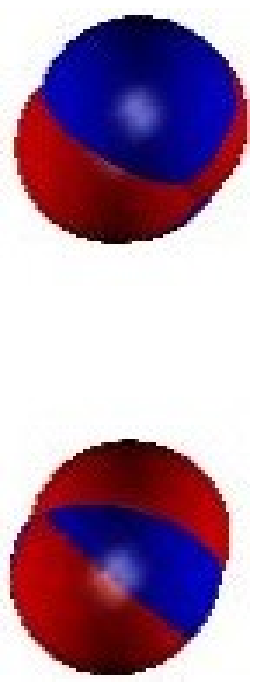}}
           \put(87,60){\includegraphics[scale=0.16]{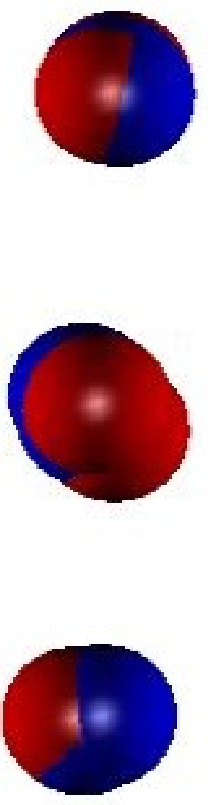}}
           \put(37.5,39.7){\includegraphics[scale=0.17]{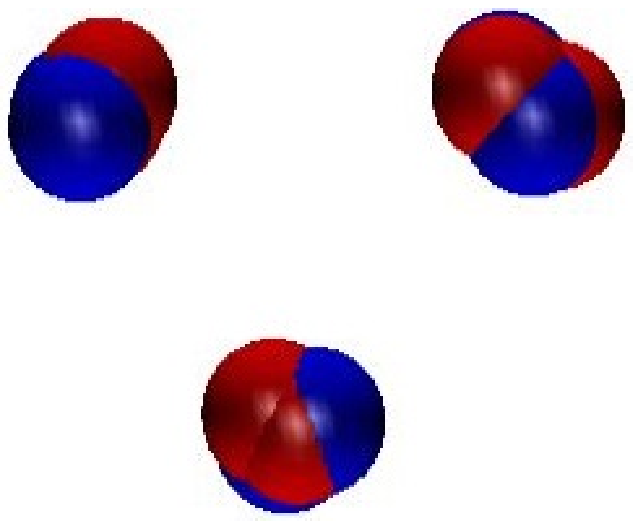}}
           \put(87,33.4){\includegraphics[scale=0.14]{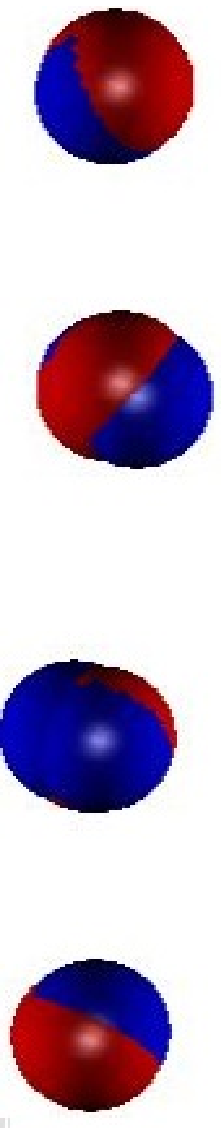}}
           \put(37,9.7){\includegraphics[scale=0.11]{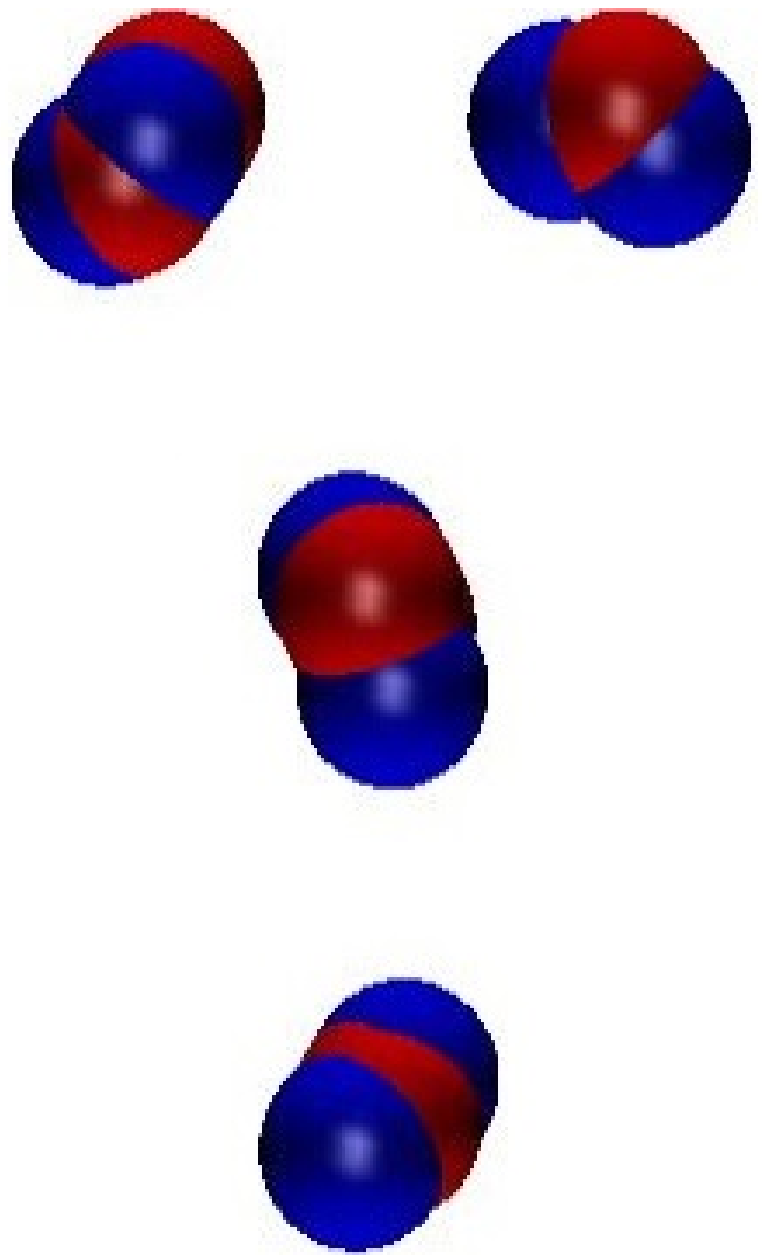}}
           \put(82,13){\includegraphics[scale=0.17]{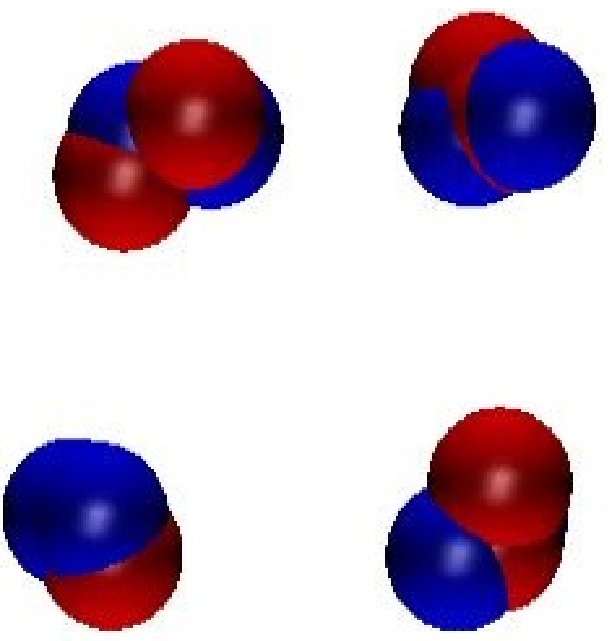}}
           \put(28,0.8){\includegraphics[scale=0.07]{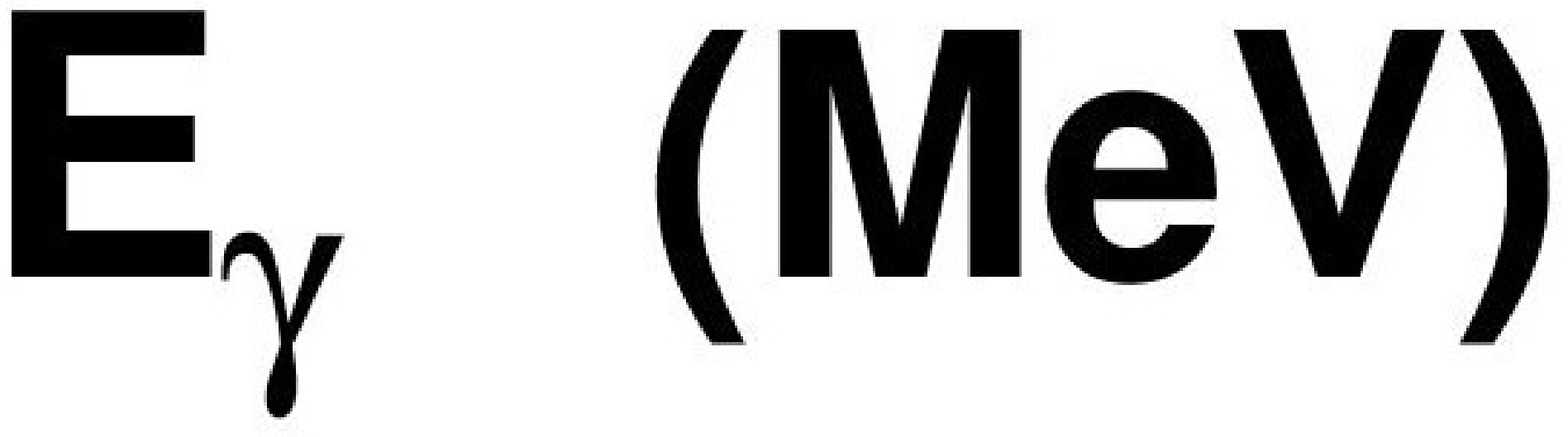}}
    \end{overpic}
\vspace{-0.7cm}
\caption{\footnotesize(color online). $^{8}$Be, $^{12}$C, and $^{16}$O GDR spectra with different cluster configurations. The corresponding $\alpha$ cluster configuration in the present EQMD model calculation is drawn in each panel, in which blue and red balls indicate protons and neutrons, respectively.The dynamical dipole evolution of  $^8$Be, $^{12}$C, and $^{16}$O with linear-chain configurations are shown in \cite{Supplemental Material}.}
\label{fig:clusterstate}
\end{figure}

For $^{12}$C, there are linear-chain and regular triangle configurations.
For $^{16}$O, we consider linear-chain, kitelike\cite{Bauhoff1984PRC29}, and square configurations.
Different configurations of $\alpha$ clustering give different  mean-field characteristics, which will essentially affect the collective motion of nucleons, e.g., in GDRs. This speculation is verified by Fig. \ref{fig:clusterstate}.

The GDR is anisotropic for $\alpha$ configurations shown in Fig. \ref{fig:clusterstate}, which originates from the fact that $\alpha$ clusters are in a plane or in a linear chain in $^{8}$Be, $^{12}$C, and $^{16}$O.
We decompose the collective motion into two directions.
One direction is perpendicular to the plane or the line of the $\alpha$ configurations,
called the short axis, indicated by long dashed red lines.
The other direction is in the plane or chain, and we take the longest axis of configuration as this direction, called the long axis, indicated by solid blue lines in Fig. \ref{fig:clusterstate}.

\begin{figure}[htbp]
\centering
\includegraphics[width=0.6\linewidth]{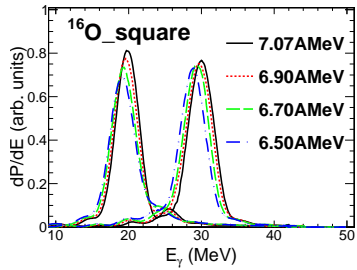}
\vspace{-0.3cm}
\caption{\footnotesize(color online). Dependence of the GDR of $^{16}$O with a square configuration on a different binding energy. The left (right) peaks originate from the GDR parallel (perpendicular) to the configuration plane.}\label{fig:square_binde}
\end{figure}

The GDR spectra along the short axis have a single peak around 30 MeV for all the cases considered in Fig. \ref{fig:clusterstate}.
It can be easily understood that the mean field along short axis is the same for different $\alpha$ cluster configurations. The peak at 30 MeV indicates the intrinsic collective dipole resonance of each $\alpha$ cluster, not affected by other degrees of freedom, which is consistent with the experimentally observed GDR of the $\alpha$-clusters in excited $A$ = 6 and 7 nuclei with possible $\alpha$ cluster structure \cite{Yamagata2004PRC69},
where each $\alpha$ feels some neighboring nucleons and, thus, has a slightly increased effective mass.

Different $\alpha$ configurations give different GDR spectra along the long axis shown by the solid blue line.
Comparing the results of $^{16}$O linear-chain [Fig. \ref{fig:clusterstate} (d)] and square [Fig. \ref{fig:clusterstate}(f)] configurations, one sees that the main peaks are at different positions, i.e., 12 MeV for linear-chain configurations and 20 MeV for square configurations.
Chain configurations with four $\alpha$'s in a chain have a larger size than four $\alpha$'s in a square configuration.
The mean field with the larger scale is responsible for a lower GDR peak, which is consistent with physics that the centroid of the GDR peak  reflects the interaction strength between clusters.

For the $^{12}$C linear-chain configuration shown in Fig. \ref{fig:clusterstate}(b), the GDR peak along the long axis has a larger width than others since it consists of two peaks, around 16 and 20 MeV.
The former peak comes from the mean field of the whole chain, and the latter peak corresponds to the two-$\alpha$-like substructure mean field, which can be confirmed by the GDR of $^{8}$Be [Fig. \ref{fig:clusterstate}(a)]. However,  no peak shows up at 20 MeV for the chain state in $^{16}$O. The reason is that there exists distructive interference between the two $^8$Be substructures in $^{16}$O.

The GDR spectrum of $^{12}$C with a triangle configuration is shown in Fig. \ref{fig:clusterstate}(c), where the strong peak around $E$ = 25 MeV can be interpreted as the coupling contribution of three-$\alpha$ clusters
which has a maximal value when the configuration is a regular triangle structure.
For the $^{12}$C triangle configuration, interactions of three-$\alpha$ clusters are superimposed and strongly affect the mean field.
The peak at $E$ = 20 MeV, originating from the two-$\alpha$ mean field, is lower than the peak which originates from the one among the three-$\alpha$ clusters.
For the $^{16}$O chain configuration and square configuration, there are also peaks located at $E$ = 25 MeV, which are consistent with the
three-$\alpha$ mean field. However, the strength is very weak.

For the kite configuration of $^{16}$O \cite{Bauhoff1984PRC29} shown in Fig. \ref{fig:clusterstate}(e), there are three peaks located at 12, 20, and 26 MeV, respectively.
The more complicated spectra are due to the reduced symmetry of the kite configuration.
In this configuration, there exists another $\alpha$ weakly bound to the three-$\alpha$ triangle structure.
The additional $\alpha$ forms a larger scale mean field than the $^{12}$C triangle configuration and gives a GDR peak located at $E$ = 12 MeV.
The peak at 20 MeV is also due to a two-$\alpha$ scale mean field.
Since there is a trianglelike substructure in the kite configuration, the peak at 26 MeV originating from the three-$\alpha$ mean-filed has larger strength, which moves from 25 MeV (in the $^{12}$C triangle system) to 26 MeV, under the influence of the weakly bound $\alpha$ cluster.
In our calculations, the excited tetrahedron (around the threshold) appears with a very small probability. In addition, this excited tetrahedron is
very unstable, which will evolve into other irregular shapes and then into
the square configuration very quickly.

In order to confirm the reliability of the results and explanations above, we study the binding-energy dependence of the GDR for the $^{16}$O square configuration, as shown in Fig. \ref{fig:square_binde}. The peaks originating from both $\alpha$ and the mean field just move a little towards the low-energy side without changing their shapes. Therefore, the number and centroid of the GDR peaks are not sensitive to the binding energy of the cluster for fixed configurations, which is consistent with the conclusion from microscopic calculations \cite{Bacca2002PRL89,Bacca2013PRL111}, namely, that the fragmented giant resonance peaks do not depend on an effective interaction.

Clusterization in light nuclei usually accompanies hyperdeformation. The very large deformation usually is measured by a rotational band, which is considered as the indirect proof of clusterization. Our calculations show that GDRs of cluster nuclei can give more detailed information about clusterization, for example, that similar GDRs of $^{8}$Be and triangle $^{12}$C appear as substructure in GDRs of chain $^{12}$C and kite $^{16}$O, respectively. Therefore, the $\alpha$ substructure can eventually be detected experimentally.

From an experimental point of view, it is feasible to obtain the expected dipole resonance described above built on $^{12}$C and $^{16}$O excited cluster states by a monochromatic Compton backscattered $\gamma$-ray beam. The GDR can be excited when the $\gamma$ energy is close  to the sum of the GDR energy and excitation energy of the cluster state;
the high excited GDR state will immediately decay to an excited cluster state and then to a ground state by emission of $\gamma$ and light particles, e.g., $\alpha$. With exclusive measurements, one can filter out other deexcitation modes and reconstruct the GDR state.
The experiment could be realized on a high intensity $\gamma$-ray source, such as the HI$\gamma$S. which has gone into operation  recently \cite{WELLER2009PPNP62}.

{\it Conclusions.}---
In summary, within a microscopic dynamical framework, we revealed how $\alpha$ configurations affect nuclear collective motion, specifically, the GDR excitation.
The dipole strength of different $\alpha$ cluster configurations have different characteristic spectra.
The characteristic spectra depicted by the number of main peaks and their centroid energies can be explained very well by the geometrical and dynamical symmetries and are insensitive to fine binding energy for given configurations.
Therefore, the GDR spectrum is a very promising unique experimental probe to study light nuclei with possible $\alpha$ cluster configurations.
The measurement of the GDR peak located around 30 MeV is a feasible way to confirm the existence of an $\alpha$ clustering state.
Analysis of other low-lying peaks can be used to diagnose the different configurations formed by $\alpha$ clusters; for example, in the GDR spectra of
chain $^{12}$C and kite $^{16}$O, there exist similar GDR spectra of $^{8}$Be and triangle $^{12}$C, respectively.

{\it Acknowledgements.}---The authors are indebted to Peter Schuck for his significant
comments, suggestions, and reading of  previous versions of the manuscript.  We thank T. Maruyama  for providing us the EQMD code and  discussions. We are grateful to  J. B. Natowitz, S. Shlomo, and R. Wada for interesting discussions and comments.
This work is partially supported by the Major State Basic Research Development Program in China under Contracts No. 2014CB845401 and 2013CB834405 and the National Natural Science Foundation of China under Contracts No. 11035009, No. 11305239, No. 11075195, and No. 11205230.

%\bibliography{ref}

\end{CJK*}
\end{document}